\PassOptionsToPackage{unicode}{hyperref}
\PassOptionsToPackage{hyphens}{url}
\documentclass[
]{article}
\usepackage{amsmath,amssymb}
\usepackage{lmodern}
\usepackage{iftex}
\ifPDFTeX
  \usepackage[T1]{fontenc}
  \usepackage[utf8]{inputenc}
  \usepackage{textcomp} 
\else 
  \usepackage{unicode-math}
  \defaultfontfeatures{Scale=MatchLowercase}
  \defaultfontfeatures[\rmfamily]{Ligatures=TeX,Scale=1}
\fi
\IfFileExists{upquote.sty}{\usepackage{upquote}}{}
\IfFileExists{microtype.sty}{
  \usepackage[]{microtype}
  \UseMicrotypeSet[protrusion]{basicmath} 
}{}
\makeatletter
\@ifundefined{KOMAClassName}{
  \IfFileExists{parskip.sty}{%
    \usepackage{parskip}
  }{
    \setlength{\parindent}{0pt}
    \setlength{\parskip}{6pt plus 2pt minus 1pt}}
}{
  \KOMAoptions{parskip=half}}
\makeatother
\usepackage{xcolor}
\usepackage{longtable,booktabs,array}
\usepackage{calc} 
\usepackage{etoolbox}
\makeatletter
\patchcmd\longtable{\par}{\if@noskipsec\mbox{}\fi\par}{}{}
\makeatother
\IfFileExists{footnotehyper.sty}{\usepackage{footnotehyper}}{\usepackage{footnote}}
\makesavenoteenv{longtable}
\usepackage{graphicx}
\makeatletter
\def\maxwidth{\ifdim\Gin@nat@width>\linewidth\linewidth\else\Gin@nat@width\fi}
\def\maxheight{\ifdim\Gin@nat@height>\textheight\textheight\else\Gin@nat@height\fi}
\makeatother
\setkeys{Gin}{width=\maxwidth,height=\maxheight,keepaspectratio}
\makeatletter
\def\fps@figure{htbp}
\makeatother
\ifLuaTeX
  \usepackage{selnolig}  
\fi
\IfFileExists{bookmark.sty}{\usepackage{bookmark}}{\usepackage{hyperref}}
\IfFileExists{xurl.sty}{\usepackage{xurl}}{} 
\hypersetup{
  hidelinks,
  pdfcreator={LaTeX via pandoc}}

\author{}
\date{}

\begin{document}

\textbf{Systematic Mapping of Monolithic Applications to Microservices Architecture}

Momil Seedat\textsuperscript{1}, Qaisar Abbas\textsuperscript{2}, Nadeem Ahmad\textsuperscript{3}

Faculty of Information Technology

University of Central Punjab  Lahore, Pakistan

\textbf{Abstract}

The aim of this paper to provide the solution microservices architecture as a popular alternative to monolithic
architecture. It discusses the advantages of microservices and the
challenges that organizations face when transitioning from a monolithic system. It presents a case study of a financial application and proposed techniques for identifying microservices on monolithic systems using domain-driven development concepts. In recent years, microservices architecture has emerged as a new architectural style in the software development industry. As legacy monolithic software becomes too large to manage, many large corporations are considering converting their traditional monolithic systems into small-scale, self-contained microservices. However, migrating from monolithic to microservices architecture is a difficult and challenging task. It presents a comparison of the two architectural styles and discusses the difficulties that led companies to switch to microservices. The study\textquotesingle s findings suggest that the proposed technique can improve work performance and establish clear models, but it may not be useful for systems with lower levels of complexity.  This research paper has practical implications for software architects and developers who are considering migrating from monolithic to microservices architecture. 


{introduction.-in-todays-world-many-organizations-run-their-businesses-using-large-enterprise-applications.-over-the-past-several-years-as-the-business-is-growing-organizations-are-confronting-problems-in-managing-and-scaling-their-applications.-in-order-to-overcome-this-problem-martin-fowler-1-defined-microservices-architecture-as-an-approach-for-developing-a-suite-of-small-services-working-as-a-single-application-where-the-services-communicate-with-lightweight-mechanisms-such-as-http.}

\section{1. Introduction}

{In today's world, many organizations run their businesses using large enterprise applications. Over the past several years, as the business has grown, Organizations are confronting problems in managing and scaling their applications. In order to overcome this problem, Martin Fowler {[}1{]} defined Microservices Architecture as an approach for developing a suite of small services working as a single application where the services
communicate with lightweight mechanisms, such as
HTTP.}

1.1. \textbf{Microservices Architecture.}  

Microservices architecture is
a distributed application where all its modules are microservices
{[}1{]}. Each service in microservices architecture (MA) has its own functionality where failure in one service does not impact the whole application. In microservices, related modules are encapsulated into a
service that provides high cohesion inwards and loose coupling outwards
{[}4{]}. This is the reason why many companies including Google, Amazon,
IBM and Netflix migrated their monolithic architecture to microservices
architecture {[}2{]}.

1.2. \textbf{Monolithic Architecture.} 

A monolithic architecture is
defined as a single code application where all components of the
application are tightly coupled with each other {[}10{]}. Monolithic
applications are easier to develop, test, and deploy {[}4{]}. Starting
Any project with this type of architecture can be a good idea. However,
When an application grows in size, it becomes harder and more strenuous
to understand and modify the system, which may result in slower
development {[}3{]}, struggle in locating errors in code, complexity in
the structure of code, and trouble with developers working in the same
environment.

1.3. \textbf{Architecture Decomposition.}

To address these issues, microservices architecture recently gained popularity in the IT world. It started as a trend in the software engineering industry as advocated by Lewis and Flower {[}4{]}. However, starting a new application with microservice architecture can be costly and time-consuming because all of its components are separately managed. Therefore, extracting small components from monolithic architecture and developing new functionality
as a microservice is a better approach. In this research article, we
proposed an innovative technique to decompose the monolithic application into microservices. We encountered several challenges during migration. Eventually, we successfully applied this technique to a real-world financial application that has more than 1.2 million users.

\section{2. \textbf{Literature Review.} }

{In most cases, the requirement for architectural amendment from monolithic to microservices is noticed when the code and the size of the company have grown up. In these cases, there are new refactoring and structure challenges to the existing system that goes along with the microservice design. It becomes difficult to maintain the monolithic system.}{2. Literature Review. In most cases, the requirement for architectural amendment from monolithic to microservices is noticed when the code and the size of the company have grown up. In these cases, there are new refactoring and structure challenges to the existing system that goes along with the microservice design. It becomes difficult to maintain the monolithic system.}

2.1. \textbf{Background and Motivation.} 

The popularity of microservices was brought to light in 2008 when a single mistake resulted in significant data corruption and prolonged outages. That's once Netflix architects set out to move the complete application from a monolithic design to AWS cloud-based microservices. The main goal of this migration was to improve accessibility, scalability, and speed {[}9{]}. They
wished Netflix to be on the market round the clock, work fast, and scale entirely. Similarly, When Uber increased its services in multiple cities, they introduced new products. The system started to grow rapidly, and that's when maintaining the monolithic architecture became an actual challenge {[}9{]}. Deploying the whole code at once, turned into hampering non-stop integration. The other disadvantage was that developers who were working for a long term at the Uber application could only make modifications to the system. One change became a large responsibility due to dependencies between the different modules of the application. So Uber decided to divide the monolith into a couple of codebases to shape a service-oriented architecture (SOA) -- or, to be more precise, a microservices architecture {[}9{]}.

2.2. \textbf{Migration of Architecture. } 

The shift from monolithic architecture to microservices has
gained popularity as companies aim to modernize their software
applications with more flexible and scalable architectures. However,
this transition poses several challenges. It requires careful planning,
communication between teams, and continuous testing to ensure that the
new architecture operates effectively. Additional challenges include
managing multiple services, handling the complexity of distributed
systems, and managing potential impacts on application performance. It
is crucial to implement proper security measures, such as securing
communication channels between services and implementing authentication
and authorization mechanisms, to ensure the safety of the application
{[}23{]}. The monolithic architecture is well-suited for small-scale
applications that have straightforward requirements. However,
microservices architecture is better suited for complex applications
that have multiple functionalities. Choosing an architecture style
depends on several factors, including project complexity, scalability
needs, and the development team\textquotesingle s experience level.
These findings highlight the importance of selecting the appropriate
architecture style to ensure that it meets the project\textquotesingle s
unique requirements {[}24{]}.

In the past years, some researchers have put their best effort to
address the issues of partitioning the application into microservices.
Richardson {[}7{]} described four decomposition Strategies for
successful migration of the application to Microservice Architecture
i.e. ``decompose by business capability'', ``Decompose by domain-driven
style sub-domain'', ``Decompose by a verb or use case'' and ``Decompose
by nouns or resources''. As per our best knowledge, the previous two are
the foremost abstract patterns that need human involvement and
decision-making.

Hippchen B. et al. presented a case study {[}8{]} on developing a
microservices based application using a domain-driven model. The authors
have demonstrated that using the existing functions could be a vital
benefit of using domain-driven development with microservices. However,
it has been discovered that separating the domain model into various
diagrams with domain views requires a significant amount of work and
expertise.

Mathai et al. {[}9{]} adopted a heterogeneous graph for mapping software
elements, i.e. programs and resources, and representing the different
relationships among them, i.e. function calls, inheritance, etc., and
employed a constraint-based clustering procedure on a novel
heterogeneous Graph Neural Network (GNN). The performed experiments
proved that the proposed approach is successful on different types of
monoliths. However, the GNN needs a pre-training phase for encoder and
decoder which can be computationally expensive.

Li et al. {[}10{]} proposed a dataflow-driven semi-automatic
decomposition approach characterized by four main steps: (i) generation
of use case and business logic specification; (ii) construction of the
fine-grained Data Flow Diagrams (DFD) and the process-datastore version
of DFD (DFDPS) representing the business logic; (iii) extraction of the
relationships between processes and data stores into decomposable
sentence sets; and (iv) clustering of processes and their closely
related datastores into single modules from the decomposable sentence
sets for the identification of potential microservices. One of the
limitations of the proposed approach is the efficiency of design.

Sellami et al. {[}11{]} considered the source code of the source
application as the input, and computed the similarities and
relationships between all the system classes from their interactions and
the domain terminology adopted within the code. Then, they used a
variant of a density-based clustering algorithm on the similarity values
to create a hierarchical structure of the candidate microservices,
together with potential outlier classes. However, such a solution is
highly dependent on the selection of the DBSCAN hyperparameters.

Brito et al. {[}12{]} proposed a static approach for detecting
microservices in a legacy software system based on topic models. They
used Latent Dirichlet Allocation (LDA) to identify the systems' topics,
corresponding to domain terms, and representing the microservices
implemented by that legacy system. A clustering method is then employed
on the graph created from the combined topics for detecting the
microservices. The limitation is that LDA needs to know the number of
topics beforehand.

Raj and Ravichandra {[}13{]} introduced an approach for extracting
microservices from a Service-Oriented Architecture (SOA) based on
graphs. More specifically, they created four procedures: (i) building
the Service Graph (SG), (ii) building the Task Graph (TG) for each
service of the SOA, (iii) detecting potential microservices using the SG
of SOA, and (iv) building a SG for a microservices application to keep
the dependencies between the generated microservices. Although the
proposed approach extracts the microservices, it is suitable for SOA
architectures instead of monoliths.

Kalia et al. {[}14{]} proposed the Mono2Micro system, an AI-based
toolchain that generates recommendations for splitting legacy web
applications into microservice partitions. Static and runtime
information are collected from a monolithic application and processed
using a tempo-spatial clustering method to produce recommendations for
splitting the application classes. Partitions are generated according to
business functionalities and data dependencies. Although the framework
is dynamic, it is only adopted for migrating legacy Java Enterprise
Edition (JEE) applications toward a microservice architecture.

In {[}15{]}, a novel method is presented based on a knowledge-graph to
assist the extraction of microservices during the initial steps of
re-architecting existing applications. Based on the method of
microservice extraction of the AKF principle, that is a well-known
principle of microservice design in the industry, the construction of
the knowledge graph is performed by designing and automatically
extracting four kinds of entities and four types of entity-entity
relationships from specification and design artifacts of the monolithic
application. A constrained Louvain algorithm is adopted to detect
microservice candidates. In the future, this work will need to introduce
more entity types to better reflect the system composition, and to split
the data of a single system at the field level to obtain more accurate
results.

Also, Dehgani {[}16{]} proposes a new approach that uses model-driven
reverse engineering (MDRE) and reinforcement learning (RL) for
supporting the migration of a monolithic code to a microservices system.
The inputs of the framework are: (i) the existing source code, (ii) the
entity-relationship (ER) model of the original system, and (iii) its use
case model. In the first step, a model of the source code of the system
is automatically extracted. Then, a model of the microservices is
derived by identifying microservices from ER and use case models.
Finally, reinforcement learning is used to propose a mapping of the
model in the first step toward the set of microservices detected in the
second step. The main limitation of the proposed approach is that a
training phase is required in the RL procedure.

Nitin et al. {[}17{]} introduced CARGO (short for Context-sensitive
lAbel pRopaGatiOn), which is a new un-/semi-supervised partition method
to refine and enrich the partitioning quality of the current
state-of-the-art algorithms employing a context- and flow-sensitive
system dependency graph of the monolithic application. Authors adopted
CARGO to augment four state-of-the-art microservice partitioning
approaches, including 1 industrial tool and 3 open-source projects.
Evaluation of the proposed approach is only performed on Java EE
applications.

The researcher L De Lauretis has provided five-step process for
migrating monolithic architecture to microservices architecture
{[}25{]}. The first step is function analysis, in which the
application\textquotesingle s functions are analyzed to identify the
business functionalities that can be isolated into microservices. The
second step is business functionalities identification, in which the
business functionalities that can be separated from the monolith and
implemented as individual microservices are identified. The third step
is business functionalities analysis, which involves analyzing each
business functionality to determine the scope of each microservice and
the communication protocols that will be used to interact with other
microservices. The fourth step is business functionalities assignment,
in which each business functionality is assigned to a microservice, and
the boundaries and responsibilities of each microservice are defined.
Finally, the fifth step is microservices creation, which involves
creating the microservices and implementing security measures such as
authentication and authorization to ensure secure communication between
microservices. These steps provide a comprehensive and structured
approach to migrating from monolithic architecture to microservices
architecture. By following these steps, organizations can effectively
manage the migration process and leverage the benefits of microservices
architecture, such as scalability, flexibility, and maintainability.

In another study {[}26{]}, four steps are proposed as a structured and
iterative approach to migrating from monolithic to microservices
architecture. The first step is analysis, in which the monolithic
application is analyzed to identify its components and dependencies. The
second step is extraction, in which each component is extracted into a
separate service while preserving dependencies. The third step is
refactoring, in which the extracted services are refactored to ensure
adherence to microservices architecture principles, such as loose
coupling and single responsibility. The fourth and final step is
orchestration, in which an orchestration layer is implemented to manage
communication between the microservices, such as through an API gateway
or service mesh. Ivanov and Tasheva have described seven steps to
decompose monolith applications into microservices {[}27{]}.

\begin{enumerate}
    \item Determine the system\textquotesingle s boundaries and identify functional areas.
    \item  Develop a business capability map and prioritize functional areas.
    \item Refine functional areas into business capabilities.
    \item Analyze interdependence between capabilities and determine service boundaries.
    \item Develop service contracts for each capability.
    \item  Develop services and orchestrate them.
    \item Test and deploy services.
\end{enumerate}

A study conducted by Zaki et al. (2022) proposed a new software
development framework for healthcare systems that uses cloud computing
and microservices architecture {[}28{]}. The framework consists of four
components: cloud platform, microservices, APIs, and containers. The
authors demonstrate the feasibility of their framework through a case
study of a healthcare system for remote patient monitoring. The
framework\textquotesingle s advantages include improved scalability,
reliability, and cost-effectiveness for healthcare applications such as
telemedicine, medical imaging, and electronic health records. Haugeland
et al. (2021) presents a migration process that enables the
transformation of monolithic applications into cloud-native
microservices-based applications {[}29{]}. The authors recommend a
gradual migration approach, starting with the most critical
functionality and adding more microservices over time. To demonstrate
the feasibility of this approach, the authors presented a case study of
a healthcare system that underwent this migration process. The new
system is a multi-tenant cloud-native application that provides
customizable healthcare solutions for different organizations. The
microservices architecture allows for greater scalability and
flexibility, while the multi-tenant model enables organizations to
customize the application to their specific needs. In conclusion, the
migration from monoliths to microservices-based customizable
multi-tenant cloud-native applications can offer several benefits to a
range of applications such as healthcare, finance, and e-commerce. In
another study, a semi-automatic tool to split a software system into
microservices, called Pangaea, was proposed {[}18{]}. It is based on the
following steps: (i) a high-level model of the system is taken in input;
(ii) decomposition is formulated as an optimization problem, and (iii) a
visual representation is adopted to map a proposed placement of
functionalities and data onto microservices, for supporting in reasoning
on the overall architecture. Pangaea makes an assessment of design
concerns, communication overheads, data management requirements,
opportunities and costs of data replication. As a future work, the
authors will need to: (i) improve the model to provide more fine-grained
modeling, (ii) to improve performance and scalability by assessing
different solving strategies, (iii) provide interactive modifications of
solutions by extending the visualization tool.

The migration process from monoliths to microservices can bring
performance benefits such as reduced latency, improved throughput, and
better resource utilization. However, there are trade-offs between
performance and modularity. The migration process can lead to increased
network overhead, more complex deployment, and reduced fault tolerance
{[}30{]}. In 2020, Santos and Silva have given a complexity metric for
migrating monolithic applications to microservices-based architecture
{[}31{]}. The metric considers the number of microservices, the
complexity of the communication between the microservices, and the
complexity of the data flow between the microservices. The authors argue
that this metric can help organizations obtain a quantitative measure of
the complexity of the migration process and allow them to assess the
potential risks and benefits of the migration.

The study conducted by the authors Blinowski, Ojdowska and Przybyłek in
2022 involves testing the performance and scalability of a monolithic
and microservice architecture using a benchmark application {[}32{]}.
The results showed that the microservice architecture outperformed the
monolithic architecture in terms of response time and throughput. The
microservice architecture was also found to be more scalable, as it was
able to handle a higher number of requests per second. However, the
authors note that implementing a microservice architecture can be more
complex and require more effort than a monolithic architecture.
Additionally, they caution that a poorly designed microservice
architecture can result in worse performance and scalability than a
well-designed monolithic architecture.

Microservice decomposition approach projected in our work is driven by
data flow diagrams from business logic. By supporting the target
operations and data extracted from the real-world application, our
approach will deliver a clear and refined architecture.

\section{3. \textbf{Technical Challenges.}}

Back in the 1990s, traditionally an internet company was supposed to run a big monolithic program on a server which is maintained due to the promise to facilitate end customers. To serve an increase in traffic a large company would simply add more instances of the
monolith.{3. Technical Challenges. Back in the 1990s, traditionally an internet company was supposed to run a big monolithic program on a server which is maintained due to the promise to facilitate end customers. To serve an increase in traffic a large company would simply add more instances of the monolith.}

A monolith centralizes the codebase so the engineers can
step through any part of the code when they are debugging the
application. Also, user requests are completely served by a monolith as it does not require many calls across a network which reduces the chances of network failures. Most software companies have their code in a monolith pattern. When these monoliths have problems, fixing them in a centralized location leads to tight couplings that are difficult to break, which can cause a variety of problems for teams working in the same environment {[}22{]}. If the program is too big it will be impossible to run it on a typical traditional machine.

3.1. \textbf{Architecture Comparison between Monolithic and
Microservices.} 

Monolithic architecture is the usual place to start when
developing an application. When creating and deploying a monolithic application, every component is included in a single unit. Figure 1 shows the components of a typical monolithic application, which includes a user interface (UI) layer, business logic layer, and data access layer that interacts with the database {[}17{]}. The major issue with monolithic applications is that it get worse in complexity when the application structure gets bigger. It becomes more difficult to integrate existing components or add new features {[}5,21{]}. As a result, new developers become more confused and the possibility of bugs and errors also increases due to uncertain situations.
\begin{center}
  \includegraphics[width=1.95833in,height=2.05764in]{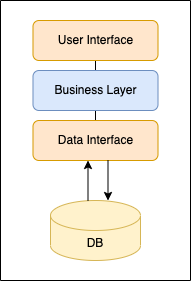}
\end{center}
\begin{center}
\textbf{Figure 1.} Typical Monolithic Application Consisting of Three
Layers.
\end{center}

Beyond the size of the enterprise applications, there are other factors to consider when breaking up a monolith application. In a microservice architecture, it is more sensible for the software development teams to take ownership of various components of the code if they are located in different areas and their communication is insufficient {[}11{]}. This
facilitates the development process because teams in various geographic
locations are no longer at the chance of changing typical software
components. If a team is responsible for a service, it is more likely
that the code will remain clean and those technical problems will be
resolved more quickly. A monolithic approach might be the best option
for small software applications where there is no need for many software
development teams {[}3{]}. If the application is small and will likely
stay where the future is predictable. In Figure 1; there are three-layer
layers that are quite popular, particularly in enterprise software
applications. As we can see, there is generally only one database,
therefore if part of the data needed to be normalized and scaled into
different databases, this type of architecture would not be able to
accommodate it. This bounds the conclusions that the teams cannot
fulfill demands to meet the growth and firm requirements. It is possible
to choose a relational database (SQL) or a non-relational database
(NoSQL) for each service when using microservices, as demonstrated in
Figure 2. Now software development teams have more possibilities in
terms of desiring their implements {[}17,22{]}.
\begin{center}
\includegraphics[width=4.09306in,height=2.07083in]{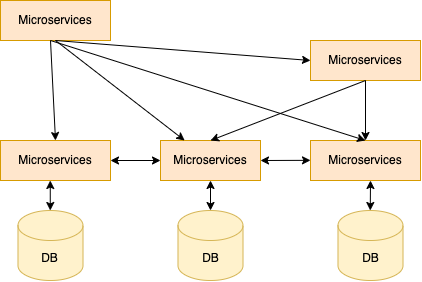}
\end{center}
\begin{center}
   \textbf{Figure 2. }Representation of Microservices Architecture
\end{center}
These two architectural types are contrasted in Table 1. As we can
notice, both architectures have their own advantages and disadvantages.
We can conclude from the table that dealing with large enterprise
applications makes the microservice architecture style more appealing.
\begin{center}
 \textbf{Table 1.} Comparing Monolithic vs Microservices~Architecture.
\end{center}
\begin{longtable}[]{@{}
  >{\raggedright\arraybackslash}p{(\columnwidth - 4\tabcolsep) * \real{0.2155}}
  >{\raggedright\arraybackslash}p{(\columnwidth - 4\tabcolsep) * \real{0.3955}}
  >{\raggedright\arraybackslash}p{(\columnwidth - 4\tabcolsep) * \real{0.3890}}@{}}
\toprule()
\begin{minipage}[b]{\linewidth}\raggedright
\textbf{Description}
\end{minipage} & \begin{minipage}[b]{\linewidth}\raggedright
\textbf{Monolithic}
\end{minipage} & \begin{minipage}[b]{\linewidth}\raggedright
\textbf{Microservices}
\end{minipage} \\
\begin{minipage}[b]{\linewidth}\raggedright
\textbf{Scalability}
\end{minipage} & \begin{minipage}[b]{\linewidth}\raggedright
Difficult to upgrade
\end{minipage} & \begin{minipage}[b]{\linewidth}\raggedright
Easy to scale per service
\end{minipage} \\
\begin{minipage}[b]{\linewidth}\raggedright
\textbf{Architecture}
\end{minipage} & \begin{minipage}[b]{\linewidth}\raggedright
Single build of a unified code
\end{minipage} & \begin{minipage}[b]{\linewidth}\raggedright
Collection of small services
\end{minipage} \\
\begin{minipage}[b]{\linewidth}\raggedright
\textbf{Programing language}
\end{minipage} & \begin{minipage}[b]{\linewidth}\raggedright
Hard to change, take a lot of time to rewrite code
\end{minipage} & \begin{minipage}[b]{\linewidth}\raggedright
Language can be selected per service
\end{minipage} \\
\begin{minipage}[b]{\linewidth}\raggedright
\textbf{Development}
\end{minipage} & \begin{minipage}[b]{\linewidth}\raggedright
Depend on the team to perform the parallel operation in same application
code
\end{minipage} & \begin{minipage}[b]{\linewidth}\raggedright
Team don't have to work parallel because each service can be delivered
independently
\end{minipage} \\
\begin{minipage}[b]{\linewidth}\raggedright
\textbf{Reliability}
\end{minipage} & \begin{minipage}[b]{\linewidth}\raggedright
If service fails, the entire application goes down
\end{minipage} & \begin{minipage}[b]{\linewidth}\raggedright
Very reliable. If service fails, the application will not go down as a
whole
\end{minipage} \\
\begin{minipage}[b]{\linewidth}\raggedright
\textbf{Maintainability}
\end{minipage} & \begin{minipage}[b]{\linewidth}\raggedright
Large code base intimidating to new developers
\end{minipage} & \begin{minipage}[b]{\linewidth}\raggedright
Small code base easier to manage
\end{minipage} \\
\begin{minipage}[b]{\linewidth}\raggedright
\textbf{Agility}
\end{minipage} & \begin{minipage}[b]{\linewidth}\raggedright
Not flexible and impossible to adopt new tech, language or frameworks
\end{minipage} & \begin{minipage}[b]{\linewidth}\raggedright
Integrate with new technologies to solve business purposes
\end{minipage} \\
\begin{minipage}[b]{\linewidth}\raggedright
\textbf{Testing}
\end{minipage} & \begin{minipage}[b]{\linewidth}\raggedright
End-to-end testing
\end{minipage} & \begin{minipage}[b]{\linewidth}\raggedright
Independent components need to be tested individually
\end{minipage} \\
\begin{minipage}[b]{\linewidth}\raggedright
\textbf{Resiliency}
\end{minipage} & \begin{minipage}[b]{\linewidth}\raggedright
One bug or issue can affect the whole system
\end{minipage} & \begin{minipage}[b]{\linewidth}\raggedright
A failure in one microservice does not affect other services
\end{minipage} \\
\begin{minipage}[b]{\linewidth}\raggedright
\textbf{Price}
\end{minipage} & \begin{minipage}[b]{\linewidth}\raggedright
Higher once the project scales
\end{minipage} & \begin{minipage}[b]{\linewidth}\raggedright
Higher at the first development stages
\end{minipage} \\
\midrule()
\endhead
\bottomrule()
\end{longtable}

\section{4. \textbf{The Proposed Migration Technique. }}

Our proposed mechanism for reducing the complexity of microservice decomposition using domain-driven design involves the following steps:

\textbf{Step I. Identification of Use Cases}

Identify the set of use cases U=\{u1, u2, u3, ...\}

Create a data flow diagram for each use case in U

\textbf{Step II. Identification of Bounded Contexts}

Identify the set of bounded contexts BC with the help of domain experts.
If a bounded context has sub-contexts, identify the inner bounded
contexts.

\textbf{Step III. Identification of Entities, Aggregates, and Domain
Services}

For each bounded context B in BC, identify the set of entities E,
aggregates A, and domain services D.

For each E, A, and D in B, identify the set of business processes BP
that use them to get data from other processes.

\textbf{Step IV. Reduction of Complexity using Combination Functions}

Define a set of systems S,( a set of entities E, a set of aggregates and
a set of domain services D). Additionally, define a set of business
processes BP.

Define the functions CombineSystems: S × BP → S\textquotesingle,
CombineProcesses: BP × T → BP\textquotesingle, and
CombineSystemsAndProcesses: \{S\} × \{BP\} → \{S\textquotesingle\}.

Identify nodes in G that share edges and can be combined into a single
node.

Use the CombineSystems and CombineProcesses functions to merge nodes
into a single node.

Update G with the new nodes and edges.

Repeat steps 3-5 until no further nodes can be combined.

Return the resulting graph as the optimized system. The result would be
the microservices.

Repeat the same process for each bounded context.

\textbf{Step V. Use of Aggregator Service}

If business use cases are independently dependent on more than 1 bounded
context, use an aggregator service that will interact with multiple
bounded contexts at the same time.

Use with caution, as it can increase the number of service calls.

\textbf{Step VI. Use of Anticorruption Layer}

Use an Anticorruption Layer if you need to interact with the older
system.

\textbf{VIII. Use of API Gateway}

Use an API Gateway to make the migration to microservices transparent to
clients.

\section{5. \textbf{Case Study.} }

This study describes the real-world adoption and
implementation of domain-driven Microservice architecture in a fintech
application. The existing system is based on monolithic architecture
where all its components are combined into a single-tiered application.
As the requirements in the system evolved, the current application
became complex and large, therefore it was difficult for the developers
to understand and make changes in the code. Often developers find it
hard to trace the exceptions in code when a system crash occurs during
peak hours. We, therefore, made the decision to convert some parts of
the application into microservices.

The case study is based on an online financial application that offers
loans to customers once they\textquotesingle ve been successfully
verified. The customer is then qualified to submit a loan request
following the verification process. The system evaluates the
customer\textquotesingle s credit score and assigns a specific limit
based on their score and desired amount. The customer can then carry out
a variety of transactions, including paying bills, transferring money,
and withdrawing cash.

5.1. \textbf{Decomposition by Domain Driven Method.} 

To decompose the
architecture, a Domain driven approach is used. Domain-driven design
(DDD) is a software development strategy involving modeling and
designing software to fit a business domain. This was initially outlined
by Eric Evans in 2003 in his book "Domain-Driven Design Tackling
Complexity in the Heart of Software" {[}18{]}.

DDD helps to solve the problem into smaller pieces. In the context of
building applications, DDD talks about problems as domains. A domain is
a business area or procedure in a company {[}11{]} {[}12{]} that
describes independent problem areas as Bounded Contexts (each Bounded
Context correlates to a microservice) {[}13{]}. Depending on the
organization, there may be a number of domains, each with its own set of
subdomains. Sub-domains are groups of related business rules and
responsibilities. So the bounded context can be used to define the
boundaries around the group of sub-domains. The sub-domains within those
boundaries have the potential to become microservices.

5.2. \textbf{Domain Analysis.} 

First, we examined the system domain by
identifying business use cases that could be converted into
microservices.

\begin{center}
\textbf{Table 2. }Application Domain Use Cases
\end{center}

\begin{longtable}[]{@{}
  >{\raggedright\arraybackslash}p{(\columnwidth - 2\tabcolsep) * \real{0.1151}}
  >{\raggedright\arraybackslash}p{(\columnwidth - 2\tabcolsep) * \real{0.8849}}@{}}
\toprule()
\begin{minipage}[b]{\linewidth}\raggedright
\textbf{\textsc{NO}}
\end{minipage} & \begin{minipage}[b]{\linewidth}\raggedright
\textbf{\textsc{USE CASES}}
\end{minipage} \\
\begin{minipage}[b]{\linewidth}\raggedright
\textbf{\textsc{1}}
\end{minipage} & \begin{minipage}[b]{\linewidth}\raggedright
Customer submit kyc for onboarding
\end{minipage} \\
\begin{minipage}[b]{\linewidth}\raggedright
\textbf{\textsc{2}}
\end{minipage} & \begin{minipage}[b]{\linewidth}\raggedright
Agent verify customer Data
\end{minipage} \\
\begin{minipage}[b]{\linewidth}\raggedright
\textbf{\textsc{3}}
\end{minipage} & \begin{minipage}[b]{\linewidth}\raggedright
Agent approve or disapprove customer
\end{minipage} \\
\begin{minipage}[b]{\linewidth}\raggedright
\textbf{\textsc{4}}
\end{minipage} & \begin{minipage}[b]{\linewidth}\raggedright
Customer submit loan request
\end{minipage} \\
\begin{minipage}[b]{\linewidth}\raggedright
\textbf{\textsc{5}}
\end{minipage} & \begin{minipage}[b]{\linewidth}\raggedright
Customer repay loan
\end{minipage} \\
\begin{minipage}[b]{\linewidth}\raggedright
\textbf{\textsc{6}}
\end{minipage} & \begin{minipage}[b]{\linewidth}\raggedright
Customer perform bill payment transaction
\end{minipage} \\
\begin{minipage}[b]{\linewidth}\raggedright
\textbf{\textsc{7}}
\end{minipage} & \begin{minipage}[b]{\linewidth}\raggedright
Customer transfer funds
\end{minipage} \\
\begin{minipage}[b]{\linewidth}\raggedright
\textbf{\textsc{8}}
\end{minipage} & \begin{minipage}[b]{\linewidth}\raggedright
Customer withdraw money from account
\end{minipage} \\
\begin{minipage}[b]{\linewidth}\raggedright
\textbf{\textsc{9}}
\end{minipage} & \begin{minipage}[b]{\linewidth}\raggedright
Customer deposit money to the account
\end{minipage} \\
\midrule()
\endhead
\bottomrule()
\end{longtable}

After defining functional requirements, a Data Flow Diagram (DFD) is
used to represent the business function of our monolithic application.
The DFD of financial application is presented in figure 3. The DFD is
composed of four elements.

\textbf{Process Notation 5.2.1}\emph{. Refers to a process that
transforms data flows. All processes must have inputs and outputs on a
DFD because they transform incoming data into outgoing data.} \emph{A
symbol circle is used to depict the process\textbf{.}}

\textbf{Datastore Notation} \textbf{5.2.2.} \emph{Represents a
collection of data items to be stored in the system.}

\textbf{Data Flow Notation 5.2.3}\emph{. Represents the direction in
which data flow and movements are represented by using arrows.}

\textbf{External Entity Notation} \textbf{5.2.4.} \emph{Demonstrates the
data flow between external systems, as well as the origin and
destination system. The external entity is represented by a rectangle.}
\begin{center}
\includegraphics[width=5.72639in,height=3.54444in]{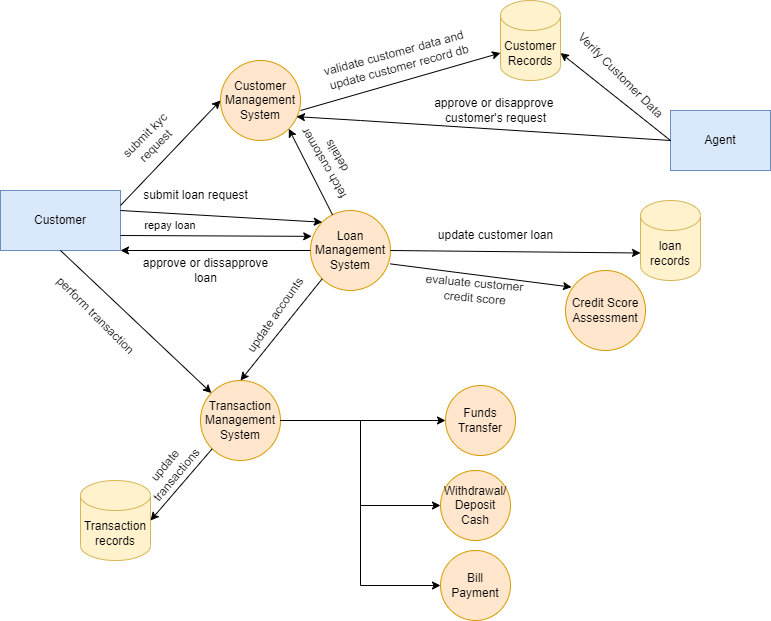}
\end{center}
\begin{center}
\textbf{Figure 3.}Data Flow Diagram of Financial Application.
\end{center}
5.3. \textbf{Identification of Bounded Context.} 

The bounded context is
a logical term that means "part of the software where particular terms,
definitions, and rules apply consistently" {[}19,20{]}. Bounded context
defines the domain boundaries, which can be determined by speaking with
domain experts. With the assistance of domain experts, we identified
three boundaries in our monolithic system i.e; customer onboarding,
loan, and transactions as shown in figure 4. The red solid line shows
the boundary line.
\begin{center}
\includegraphics[width=5.71042in,height=3.39167in]{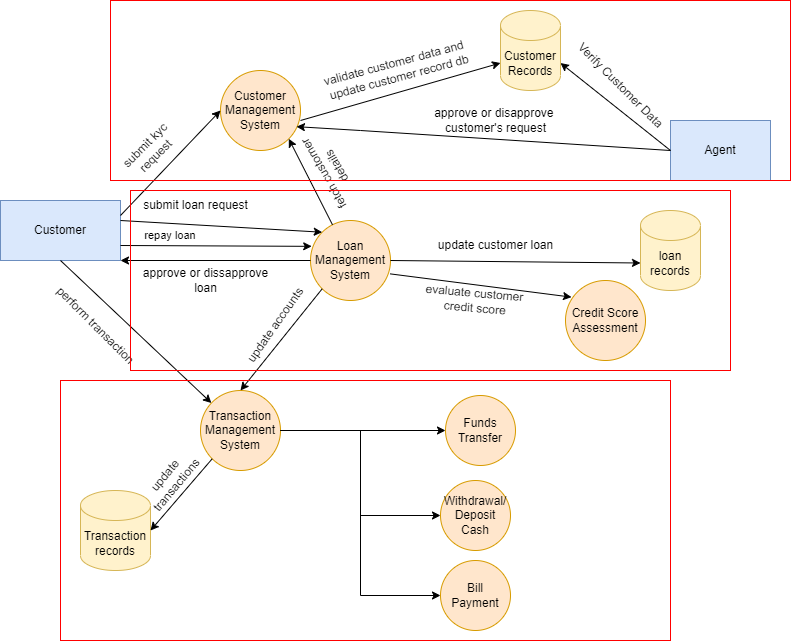}
\end{center}
\begin{center}
\textbf{Figure 4.} Identification of Bounded Context.
\end{center}
5.4. \textbf{Define Aggregates, Entities, and Domain Service.} 

The next
step after the identification is to classify aggregates, entities and
domain services in each bounded context

5.4.1. \textbf{Entities and Aggregates.} An entity is a persistent
object with a distinct identity. An entity\textquotesingle s attributes
can change over time. For example, a person\textquotesingle s address
may change, but he/she will remain the same person {[}13{]}. Aggregates
are defined as a collection of related objects treated as a single
entity {[}12,19{]}. A consistency boundary is defined by an aggregate
around one or more entities. The root of an aggregate is always the same
entity.

5.4.2. \textbf{Domain Services.} Evan described the domain service as
``A standalone operation within the context of your domain. A Service
Object collects one or more services into an object'' {[}18{]}. Domain
services assist in the modeling of behavior that includes multiple
entities. Following the identification of bounded contexts, the next
step is to identify entities, aggregates, and domain services within
each bounded context. The detailed description of entities and business
functionalities cannot be discussed due to security reasons but here is
a brief concept that how we created services.

\begin{itemize}
\item \textbf{Customers Onboard Bounded Context.} Customer and Agent are
  entities that represent onboard boundaries, whereas approval is a
  child entity of the agent that approves or disapproves the status of
  the customer. An onboarding service is identified as a domain service
  that performs business operations such as validating customer data,
  assigning requests to agents, and so on.
\item
  \textbf{Loan Bounded Context.} In a loan-bound context, customers can
  choose loan plans and submit loan requests. The loan management system
  evaluates the customer\textquotesingle s credit score and decides
  whether or not to disburse the loan. It also settles loans after they
  are repaid. The entities and aggregates identified in the following
  context are loans, loan plans, loan repayments, and loan disbursement.
  Credit assessment is a domain service that evaluates a
  customer\textquotesingle s score based on their history.
\item
  \textbf{Transactional Bounded Context.} Loan repayment, funds
  transfer, bill payment, and cash withdrawal/deposit are all aggregates
  in a transactional bounded context. A transaction service is a domain
  service that routes requests to different processes and updates
  records after each transaction. The figure 5 shows identified sets of
  entities, aggregates, and domain services.
\end{itemize}
\begin{center}
\includegraphics[width=5.49097in,height=2.58333in]{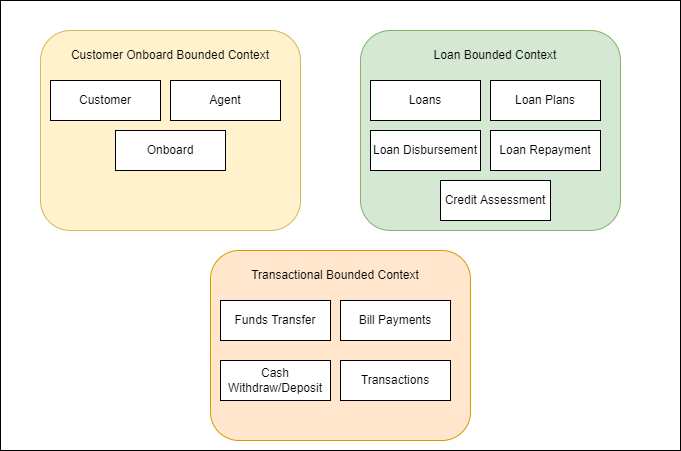}
\end{center}
\begin{center}
\textbf{Figure 5.} Entities, Aggregates, and Domain Services in
Financial Application
\end{center}
5.5. \textbf{Reduction of Complexity using Combination Functions}

Combination Function can be used to simplify complex systems by
identifying common functionalities and reducing the number of
inter-system calls. The process begins by examining systems that are
dependent on other processes and comparing them with other systems to
identify similar processes. These processes are then combined to
streamline the overall system. Additionally, processes of systems that
interact with the same database tables are compared and combined,
resulting in automatic integration of these systems.

\begin{quote}
\textbf{{Table} 3.} Comparing System Processes in Customer
bounded context
\end{quote}

\begin{longtable}[]{@{}
  >{\raggedright\arraybackslash}p{(\columnwidth - 6\tabcolsep) * \real{0.2500}}
  >{\raggedright\arraybackslash}p{(\columnwidth - 6\tabcolsep) * \real{0.2500}}
  >{\raggedright\arraybackslash}p{(\columnwidth - 6\tabcolsep) * \real{0.2500}}
  >{\raggedright\arraybackslash}p{(\columnwidth - 6\tabcolsep) * \real{0.2500}}@{}}
\toprule()
\begin{minipage}[b]{\linewidth}\raggedright
\textbf{Customer}
\end{minipage} & \begin{minipage}[b]{\linewidth}\raggedright
\textbf{Onboard}
\end{minipage} & \begin{minipage}[b]{\linewidth}\raggedright
\textbf{Agent Surveillance}
\end{minipage} & \begin{minipage}[b]{\linewidth}\raggedright
\textbf{Agent}
\end{minipage} \\
\begin{minipage}[b]{\linewidth}\raggedright
CRP
\end{minipage} & \begin{minipage}[b]{\linewidth}\raggedright
OIB
\end{minipage} & \begin{minipage}[b]{\linewidth}\raggedright
AIS
\end{minipage} & \begin{minipage}[b]{\linewidth}\raggedright
AI
\end{minipage} \\
\begin{minipage}[b]{\linewidth}\raggedright
CNCP
\end{minipage} & \begin{minipage}[b]{\linewidth}\raggedright
OBS
\end{minipage} & \begin{minipage}[b]{\linewidth}\raggedright
AAI
\end{minipage} & \begin{minipage}[b]{\linewidth}\raggedright
AIS
\end{minipage} \\
\begin{minipage}[b]{\linewidth}\raggedright
CTPO
\end{minipage} & \begin{minipage}[b]{\linewidth}\raggedright
PRR
\end{minipage} & \begin{minipage}[b]{\linewidth}\raggedright
AAPR
\end{minipage} & \begin{minipage}[b]{\linewidth}\raggedright
APC
\end{minipage} \\
\begin{minipage}[b]{\linewidth}\raggedright
\end{minipage} & \begin{minipage}[b]{\linewidth}\raggedright
\end{minipage} & \begin{minipage}[b]{\linewidth}\raggedright
\end{minipage} & \begin{minipage}[b]{\linewidth}\raggedright
AAPR
\end{minipage} \\
\midrule()
\endhead
\bottomrule()
\end{longtable}

Set of systems that were founded in customer bounded context are
(Customer, Onboard, Agent Surveillance and Agent). In order to integrate
system with business processes, for the customer bounded context, it has
been observed in Table 3 that the processes AIS and AAPR are common
between the systems Agent Surveillance and Agent. Therefore, these
systems would be combine.

\begin{quote}
\textbf{Table 4.} Comparing Processes Tables in Customer
bounded context
\end{quote}

\begin{longtable}[]{@{}
  >{\raggedright\arraybackslash}p{(\columnwidth - 20\tabcolsep) * \real{0.0909}}
  >{\raggedright\arraybackslash}p{(\columnwidth - 20\tabcolsep) * \real{0.0909}}
  >{\raggedright\arraybackslash}p{(\columnwidth - 20\tabcolsep) * \real{0.0909}}
  >{\raggedright\arraybackslash}p{(\columnwidth - 20\tabcolsep) * \real{0.0909}}
  >{\raggedright\arraybackslash}p{(\columnwidth - 20\tabcolsep) * \real{0.0909}}
  >{\raggedright\arraybackslash}p{(\columnwidth - 20\tabcolsep) * \real{0.0909}}
  >{\raggedright\arraybackslash}p{(\columnwidth - 20\tabcolsep) * \real{0.0909}}
  >{\raggedright\arraybackslash}p{(\columnwidth - 20\tabcolsep) * \real{0.0909}}
  >{\raggedright\arraybackslash}p{(\columnwidth - 20\tabcolsep) * \real{0.0909}}
  >{\raggedright\arraybackslash}p{(\columnwidth - 20\tabcolsep) * \real{0.0909}}
  >{\raggedright\arraybackslash}p{(\columnwidth - 20\tabcolsep) * \real{0.0909}}@{}}
\toprule()
\begin{minipage}[b]{\linewidth}\raggedright
\textbf{CRP}
\end{minipage} & \begin{minipage}[b]{\linewidth}\raggedright
\textbf{CNCP}
\end{minipage} & \begin{minipage}[b]{\linewidth}\raggedright
\textbf{CTPO}
\end{minipage} & \begin{minipage}[b]{\linewidth}\raggedright
\textbf{OIB}
\end{minipage} & \begin{minipage}[b]{\linewidth}\raggedright
\textbf{OBS}
\end{minipage} & \begin{minipage}[b]{\linewidth}\raggedright
\textbf{PRR}
\end{minipage} & \begin{minipage}[b]{\linewidth}\raggedright
\textbf{AIS}
\end{minipage} & \begin{minipage}[b]{\linewidth}\raggedright
\textbf{AAI}
\end{minipage} & \begin{minipage}[b]{\linewidth}\raggedright
\textbf{AAPR}
\end{minipage} & \begin{minipage}[b]{\linewidth}\raggedright
\textbf{AI}
\end{minipage} & \begin{minipage}[b]{\linewidth}\raggedright
\textbf{APC}
\end{minipage} \\
\begin{minipage}[b]{\linewidth}\raggedright
TCI
\end{minipage} & \begin{minipage}[b]{\linewidth}\raggedright
TCPA
\end{minipage} & \begin{minipage}[b]{\linewidth}\raggedright
TCPI
\end{minipage} & \begin{minipage}[b]{\linewidth}\raggedright
TOCB
\end{minipage} & \begin{minipage}[b]{\linewidth}\raggedright
TCVB
\end{minipage} & \begin{minipage}[b]{\linewidth}\raggedright
TCM
\end{minipage} & \begin{minipage}[b]{\linewidth}\raggedright
TACM
\end{minipage} & \begin{minipage}[b]{\linewidth}\raggedright
TAA
\end{minipage} & \begin{minipage}[b]{\linewidth}\raggedright
TACP
\end{minipage} & \begin{minipage}[b]{\linewidth}\raggedright
TAV
\end{minipage} & \begin{minipage}[b]{\linewidth}\raggedright
TAS
\end{minipage} \\
\begin{minipage}[b]{\linewidth}\raggedright
TCA
\end{minipage} & \begin{minipage}[b]{\linewidth}\raggedright
\end{minipage} & \begin{minipage}[b]{\linewidth}\raggedright
\end{minipage} & \begin{minipage}[b]{\linewidth}\raggedright
\end{minipage} & \begin{minipage}[b]{\linewidth}\raggedright
TCM
\end{minipage} & \begin{minipage}[b]{\linewidth}\raggedright
\end{minipage} & \begin{minipage}[b]{\linewidth}\raggedright
\end{minipage} & \begin{minipage}[b]{\linewidth}\raggedright
TAPR
\end{minipage} & \begin{minipage}[b]{\linewidth}\raggedright
TRP
\end{minipage} & \begin{minipage}[b]{\linewidth}\raggedright
TAP
\end{minipage} & \begin{minipage}[b]{\linewidth}\raggedright
TAP
\end{minipage} \\
\midrule()
\endhead
\bottomrule()
\end{longtable}

Table 4 displays the database tables utilized in each process. After
analyzing the Table 4, it is observed that the processes (OBS,PRR) share
a common table, TCM, whereas (AI,APC) have a common table, TAB. However,
it is worth noting that (OBS, PRR) and (AI,APC) already belong to the
same system. Therefore, no other system can be identified that shares
common tables that could be combined with the existing systems.

The combination of the Agent Surveillance and Agent systems is referred
to as a support system.

\begin{quote}
\textbf{Table 5.} Comparing Systems Processes in Loan bounded
context
\end{quote}

\begin{longtable}[]{@{}
  >{\raggedright\arraybackslash}p{(\columnwidth - 10\tabcolsep) * \real{0.1667}}
  >{\raggedright\arraybackslash}p{(\columnwidth - 10\tabcolsep) * \real{0.1667}}
  >{\raggedright\arraybackslash}p{(\columnwidth - 10\tabcolsep) * \real{0.1667}}
  >{\raggedright\arraybackslash}p{(\columnwidth - 10\tabcolsep) * \real{0.1667}}
  >{\raggedright\arraybackslash}p{(\columnwidth - 10\tabcolsep) * \real{0.1667}}
  >{\raggedright\arraybackslash}p{(\columnwidth - 10\tabcolsep) * \real{0.1667}}@{}}
\toprule()
\begin{minipage}[b]{\linewidth}\raggedright
\textbf{Loans}
\end{minipage} & \begin{minipage}[b]{\linewidth}\raggedright
\textbf{Loan Plans}
\end{minipage} & \begin{minipage}[b]{\linewidth}\raggedright
\textbf{Loan Disbursement}
\end{minipage} & \begin{minipage}[b]{\linewidth}\raggedright
\textbf{Loan Repayment}
\end{minipage} & \begin{minipage}[b]{\linewidth}\raggedright
\textbf{Credit Assessment}
\end{minipage} & \begin{minipage}[b]{\linewidth}\raggedright
\textbf{Risk Assessment}
\end{minipage} \\
\begin{minipage}[b]{\linewidth}\raggedright
LC
\end{minipage} & \begin{minipage}[b]{\linewidth}\raggedright
LIP
\end{minipage} & \begin{minipage}[b]{\linewidth}\raggedright
LDV
\end{minipage} & \begin{minipage}[b]{\linewidth}\raggedright
LRM
\end{minipage} & \begin{minipage}[b]{\linewidth}\raggedright
CAC
\end{minipage} & \begin{minipage}[b]{\linewidth}\raggedright
CARC
\end{minipage} \\
\begin{minipage}[b]{\linewidth}\raggedright
LV
\end{minipage} & \begin{minipage}[b]{\linewidth}\raggedright
LPP
\end{minipage} & \begin{minipage}[b]{\linewidth}\raggedright
LDI
\end{minipage} & \begin{minipage}[b]{\linewidth}\raggedright
\end{minipage} & \begin{minipage}[b]{\linewidth}\raggedright
CARC
\end{minipage} & \begin{minipage}[b]{\linewidth}\raggedright
CARIP
\end{minipage} \\
\begin{minipage}[b]{\linewidth}\raggedright
\end{minipage} & \begin{minipage}[b]{\linewidth}\raggedright
\end{minipage} & \begin{minipage}[b]{\linewidth}\raggedright
\end{minipage} & \begin{minipage}[b]{\linewidth}\raggedright
\end{minipage} & \begin{minipage}[b]{\linewidth}\raggedright
CAS
\end{minipage} & \begin{minipage}[b]{\linewidth}\raggedright
\end{minipage} \\
\midrule()
\endhead
\bottomrule()
\end{longtable}

In Table 5, it is noted that the Credit Assessment and Risk Assessment
systems share a common process known as CARC. Therefore, these systems
will be integrated

\begin{quote}
\textbf{Table 6\textbf{.}} Comparing Processes Tables in Loan bounded
context
\end{quote}

\begin{longtable}[]{@{}
  >{\raggedright\arraybackslash}p{(\columnwidth - 20\tabcolsep) * \real{0.0909}}
  >{\raggedright\arraybackslash}p{(\columnwidth - 20\tabcolsep) * \real{0.0909}}
  >{\raggedright\arraybackslash}p{(\columnwidth - 20\tabcolsep) * \real{0.0909}}
  >{\raggedright\arraybackslash}p{(\columnwidth - 20\tabcolsep) * \real{0.0909}}
  >{\raggedright\arraybackslash}p{(\columnwidth - 20\tabcolsep) * \real{0.0909}}
  >{\raggedright\arraybackslash}p{(\columnwidth - 20\tabcolsep) * \real{0.0909}}
  >{\raggedright\arraybackslash}p{(\columnwidth - 20\tabcolsep) * \real{0.0909}}
  >{\raggedright\arraybackslash}p{(\columnwidth - 20\tabcolsep) * \real{0.0909}}
  >{\raggedright\arraybackslash}p{(\columnwidth - 20\tabcolsep) * \real{0.0909}}
  >{\raggedright\arraybackslash}p{(\columnwidth - 20\tabcolsep) * \real{0.0909}}
  >{\raggedright\arraybackslash}p{(\columnwidth - 20\tabcolsep) * \real{0.0909}}@{}}
\toprule()
\begin{minipage}[b]{\linewidth}\raggedright
\textbf{LC}
\end{minipage} & \begin{minipage}[b]{\linewidth}\raggedright
\textbf{LV}
\end{minipage} & \begin{minipage}[b]{\linewidth}\raggedright
\textbf{LIP}
\end{minipage} & \begin{minipage}[b]{\linewidth}\raggedright
\textbf{LPP}
\end{minipage} & \begin{minipage}[b]{\linewidth}\raggedright
\textbf{LDV}
\end{minipage} & \begin{minipage}[b]{\linewidth}\raggedright
\textbf{LDI}
\end{minipage} & \begin{minipage}[b]{\linewidth}\raggedright
\textbf{LRM}
\end{minipage} & \begin{minipage}[b]{\linewidth}\raggedright
\textbf{CAC}
\end{minipage} & \begin{minipage}[b]{\linewidth}\raggedright
\textbf{CARC}
\end{minipage} & \begin{minipage}[b]{\linewidth}\raggedright
\textbf{CAS}
\end{minipage} & \begin{minipage}[b]{\linewidth}\raggedright
\textbf{CARIP}
\end{minipage} \\
\begin{minipage}[b]{\linewidth}\raggedright
TLC
\end{minipage} & \begin{minipage}[b]{\linewidth}\raggedright
TVG
\end{minipage} & \begin{minipage}[b]{\linewidth}\raggedright
TLI
\end{minipage} & \begin{minipage}[b]{\linewidth}\raggedright
TLP
\end{minipage} & \begin{minipage}[b]{\linewidth}\raggedright
TDD
\end{minipage} & \begin{minipage}[b]{\linewidth}\raggedright
TLDI
\end{minipage} & \begin{minipage}[b]{\linewidth}\raggedright
TLRY
\end{minipage} & \begin{minipage}[b]{\linewidth}\raggedright
TCIC
\end{minipage} & \begin{minipage}[b]{\linewidth}\raggedright
TCAR
\end{minipage} & \begin{minipage}[b]{\linewidth}\raggedright
TCIC
\end{minipage} & \begin{minipage}[b]{\linewidth}\raggedright
TCRP
\end{minipage} \\
\begin{minipage}[b]{\linewidth}\raggedright
\end{minipage} & \begin{minipage}[b]{\linewidth}\raggedright
TLI
\end{minipage} & \begin{minipage}[b]{\linewidth}\raggedright
\end{minipage} & \begin{minipage}[b]{\linewidth}\raggedright
TLC
\end{minipage} & \begin{minipage}[b]{\linewidth}\raggedright
\end{minipage} & \begin{minipage}[b]{\linewidth}\raggedright
\end{minipage} & \begin{minipage}[b]{\linewidth}\raggedright
TLLM
\end{minipage} & \begin{minipage}[b]{\linewidth}\raggedright
TCS
\end{minipage} & \begin{minipage}[b]{\linewidth}\raggedright
\end{minipage} & \begin{minipage}[b]{\linewidth}\raggedright
TCA
\end{minipage} & \begin{minipage}[b]{\linewidth}\raggedright
\end{minipage} \\
\midrule()
\endhead
\bottomrule()
\end{longtable}

According to Table 6, the processes (LC,LPP) share a common table TLI,
while the processes (LV,LIP) share a common table TLC. However, these
processes belong to different systems and were not combined in previous
versions of the system. As a result, the Loans and Loan Plans systems,
which contain these processes, will be integrated in order to improve
operational efficiency.

It should be noted that the processes (CAC,CAS) also share a common
table TCIC, as highlighted in yellow in Table 6. However, these
processes already belong to the same system and therefore do not need to
be compared.

After comparing systems and processes of the Transactional Bounded
Context, no common systems or processes were identified. Therefore, the
systems that were identified within this context, namely entities,
aggregates, and domain services, will be considered as potential
microservice candidates for this particular bounded context.

5.6. \textbf{Identification of Microservices.} After performing a
comparison of the functionality of the systems, the following systems
have been identified as microservices based on their respective contexts
In order to ensure these domains are completely dedicated to entire
teams, services were combined in a new modified domain. The identified
microservices in each bounded context are listed below.

\begin{quote}
\textbf{Table 7\textbf{.}} Identified microservices in bounded context
\end{quote}

\begin{longtable}[]{@{}
  >{\raggedright\arraybackslash}p{(\columnwidth - 4\tabcolsep) * \real{0.3234}}
  >{\raggedright\arraybackslash}p{(\columnwidth - 4\tabcolsep) * \real{0.3242}}
  >{\raggedright\arraybackslash}p{(\columnwidth - 4\tabcolsep) * \real{0.3524}}@{}}
\toprule()
\begin{minipage}[b]{\linewidth}\raggedright
\textbf{Onboard}
\end{minipage} & \begin{minipage}[b]{\linewidth}\raggedright
\textbf{Loans}
\end{minipage} & \begin{minipage}[b]{\linewidth}\raggedright
\textbf{Transactions}
\end{minipage} \\
\begin{minipage}[b]{\linewidth}\raggedright
\textbf{Customer Management Service}
\end{minipage} & \begin{minipage}[b]{\linewidth}\raggedright
Loan Management Service
\end{minipage} & \begin{minipage}[b]{\linewidth}\raggedright
Transaction Management Service
\end{minipage} \\
\begin{minipage}[b]{\linewidth}\raggedright
\textbf{Support Service}
\end{minipage} & \begin{minipage}[b]{\linewidth}\raggedright
Loan Repayment Service
\end{minipage} & \begin{minipage}[b]{\linewidth}\raggedright
Bill Payment Service
\end{minipage} \\
\begin{minipage}[b]{\linewidth}\raggedright
\textbf{Customer Onboard Service}
\end{minipage} & \begin{minipage}[b]{\linewidth}\raggedright
Loan Disbursement Service
\end{minipage} & \begin{minipage}[b]{\linewidth}\raggedright
Funds Transfer Service
\end{minipage} \\
\begin{minipage}[b]{\linewidth}\raggedright
\end{minipage} & \begin{minipage}[b]{\linewidth}\raggedright
Credit Assessment Service
\end{minipage} & \begin{minipage}[b]{\linewidth}\raggedright
Cash Withdraw and Deposit Service
\end{minipage} \\
\midrule()
\endhead
\bottomrule()
\end{longtable}

In figure 6, the rounded white rectangles in the yellow, green, and
orange colors represent the services inside the bounded context.
Additional services that are not part of any bounded context are also
represented with white rounded rectangle in Fig 6. A detailed
description of services is given below.

\begin{enumerate}
\def\labelenumi{\roman{enumi})}
\item
  \textbf{Aggregation Service.} A new service is created that receives a
  request from a customer, calls other services to combine their
  responses, and then sends the combined response back to the customers.
  For example, a customer needs to see his loan details and loan
  transactions so this service will invoke both loan and transaction and
  send the aggregated response back to the customer.
\item
  \textbf{Gateway Routing.} Requests are routed to multiple
  microservices via a single endpoint, eliminating the need for the
  client application to manage multiple endpoints. As an example, if any
  internal service is modified, the client does not always need to be
  updated. It can keep sending requests to the gateway; only the routing
  will change.
\item
  \textbf{Anti-Corruption Layer.} As we shifted some legacy system
  features to microservices. It\textquotesingle s possible that the new
  system will require legacy system resources. To allow the new services
  to call the legacy system, an anti-corruption layer is added in Fig 6
  that translates communication between the two systems.
\end{enumerate}
\begin{center}
\includegraphics[width=6.49722in,height=3.38472in]{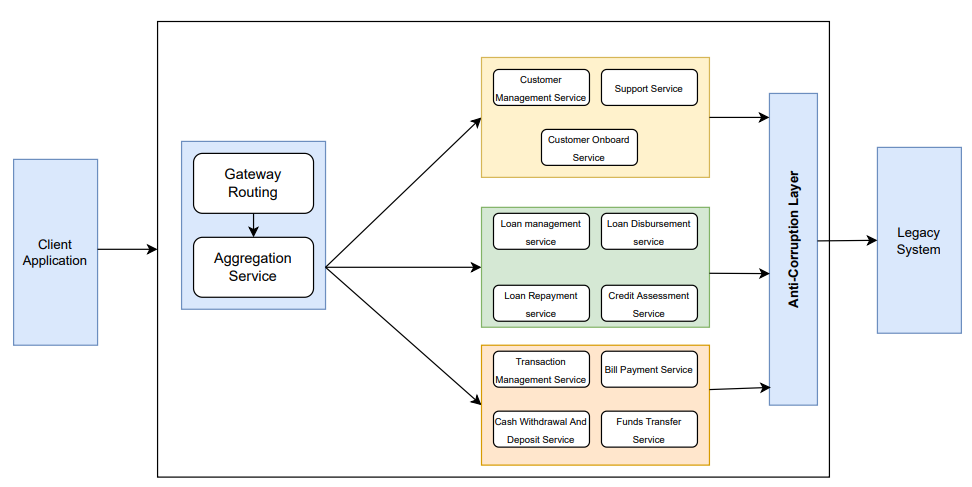}
\end{center}
\begin{center}
\textbf{{\textbf{Figure}} 6}. Migration Results of Monolithic to Microservice
Architecture
\end{center}

5.7. \textbf{Communication.} As microservices are distributed and
interact with other services on a network level, communication has
proven to be the biggest challenge. The chatty communication between
services can cause network overhead. Communication protocols like HTTP,
gRPC (modern framework based on Remote Procedure Call), or AMQP are the
only ones that the services can interact with.

HTTP and gRPC protocols are used to enable synchronized communication
between services. Since HTTP calls between microservices are relatively
simple to implement and block the operation until a result is returned
or the request time is out, and gRPC is a binary framing protocol for
data transport - unlike HTTP 1.1 {[}15{]}. gRPC is a lightweight and
highly performant protocol that can be up to 8x faster than JSON
serialization while producing messages that are 60 to 80\% smaller
{[}16{]}. RESTful APIs over HTTP protocol were decided to use for calls
from client applications because payloads were readable in RESTful API,
but for inter-service communication, we had to consider network
performance because direct HTTP calls to multiple microservices can
increase latency and degrade performance {[}15{]}. Due to this, gRPC
protocol which is significantly faster than HTTP {[}16{]} was used to
call internal systems over binary network protocols.

Through an event bus, asynchronous communication is established between
a loan management system and a transaction management system. Upon loan
settlement or disbursement, the loan service triggers an event. The
transaction service monitors these events and updates the transaction
table accordingly.
\begin{center}
\begin{quote}
\includegraphics[width=5.80208in,height=2.52639in]{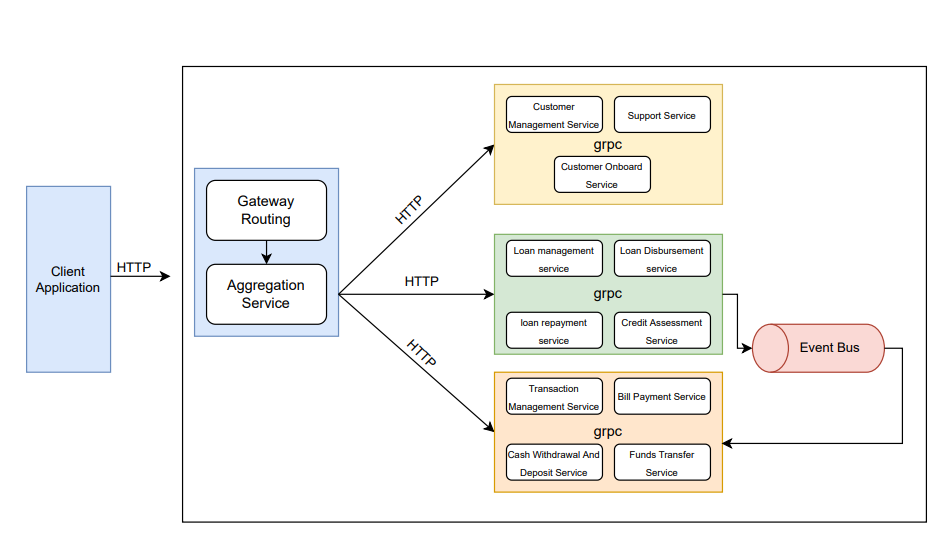}
\end{quote}
\end{center}
\begin{center}
\textbf{{Figure} 7}. Communication Protocols for Data Flow Among Services
\end{center}
6. \textbf{Summary and Findings.}

In this study, a domain-driven design
approach is used to convert a real-time monolithic application to
microservices. As today\textquotesingle s business environments are
extremely complex where any wrong moves can have disastrous
consequences. Domain Driven Designs (DDD) are able to solve our complex
business functions and assist us in defining clear domain models that
could later be easily divided into small services or subsystems. Many
developers claim that after implementing this approach, communication
among teams has improved throughout the development cycle because each
team is solely focused on their domain and each
developer\textquotesingle s roles are clearly defined. However, the
disadvantage of this approach is that it necessitates collaboration
between technical and domain experts in order to create an application
model for solving domain problems. Applications with high technical
complexity can also be challenging for business domain experts which may
lead to numerous restrictions that not all team members may be able to
overcome.

7. \textbf{Conclusions.} 

Microservices are a viable method of dividing a
large application into self-contained services. This paper described a
technique for converting monolithic application features into
microservices that was applied to a real-time application. The migration
strategy was primarily based on domain-driven design concepts, which
include the following steps: domain analysis using Data Flow Diagram,
bounded context identification, aggregates, events, and domain service
selection, as well as microservice identification. Furthermore, the
approach to service communication is also discussed in order to improve
application performance. Apart from that, we discovered that application
decomposition takes time and requires expert advice. This approach may
not be appropriate for applications with minor complexities. In the
future, we intend to convert other parts of applications in different
domains from monolithic to microservices.

\hypertarget{declarations}{%
\subsection{\texorpdfstring{\textbf{Declarations}}{Declarations}}\label{declarations}}

\textbf{Funding:} The authors did not receive support from any
organization for the submitted work. The project received no funding and
is carried out entirely on a volunteer basis by university researchers.

\textbf{Conflict of interest:} Authors have no financial and conflicting
interests that are relevant to the content of this paper.

\textbf{Acknowledgment.} The authors gratefully acknowledge the helpful
comments and suggestions of the reviewers, which have improved the
presentation.

\textbf{REFERENCES}

\begin{enumerate}
\item Dragoni, N., Giallorenzo, S., Lafuente, A. L., Mazzara, M., Montesi, F., Mustafin, R., \& Safina, L. (2017). Microservices: Yesterday, Today, and Tomorrow.
\item SmartBear Software. (2015, December 8). Why You Can't Talk About Microservices Without Mentioning Netflix. SmartBear.Com.https://smartbear.com/blog/why-you-cant-talk about-microservices-without-ment/\#:\%7E:text=A\%20well\ 2Ddesigned\%20microservices\%20architecture,up\%20with\%20its\%20growth\%20rate
\item Ponce, F., Márquez, G., \& Astudillo, H. (2019, November). Migrating from monolithic architecture to microservices: A Rapid Review. In 2019 38th International Conference of the Chilean Computer Science Society  (SCCC) (pp. 1-7). IEEE.
\item  Fritzsch, J., Bogner, J., Zimmermann, A., \& Wagner, S. (2018, March). From monolith to microservices: A classification of refactoring approaches. In International Workshop on Software Engineering Aspects of Continuous Development and New Paradigms of Software Production and Deployment (pp. 128-141). Springer, Cham.
\item Richardson,C.(2017,March).Monolithic Architecture pattern. https://microservices.io/patterns/monolithic.html
\item Axelsson, E., \& Karlkvist, E. (2019). Extracting Microservices from a
  Monolithic Application.
\item
  C. Richardson, ``Pattern: Microservice architecture,'' 22 June 2017.
  {[}Online{]}. Available:
  http://microservices.io/patterns/microservices.html
\item
  Hippchen, B., Giessler, P., Steinegger, R., Schneider, M., \& Abeck,
  S. (2017). Designing microservice-based applications by using a
  domain-driven design approach. International Journal on Advances in
  Software, 10(3\&4), 432-445. A. Balalaie, A. Heydarnoori, P. Jamshidi,
  \emph{Microservices Architecture Enables Devops}, IEEE Software, vol
  33, no. 3, 2016, pp. 42-52.
\item
  Rud, A. (2020, June 23). Why and How Netflix, Amazon, and Uber
  Migrated to Microservices: Learn from Their Experience--HYSEnterprise.
  https://www.hys-enterprise.com/blog/why-and-how-netflix-amazon-and-uber-migrated-to-microservices-learn-from-their-experience/
\item
  Kalske, M., Mäkitalo, N., \& Mikkonen, T. (2017, June). Challenges
  when moving from monolith to microservice architecture. In
  International Conference on Web Engineering (pp. 32-47). Springer,
  Cham.
\item
  S. Newman, Building Microservices, 1005 Gravenstein Highway North,
  Sebastopol, CA 95472: O'Reilly Media, Inc., 2021.
\item
  Languric, M., \& Zaki, L. (2022). Migrating monolithic system to
  domain-driven microservices\,: Developing a generalized migration
  strategy for an architecture built on microservices (Dissertation).
  Retrieved from http://urn.kb.se/resolve?urn=urn:nbn:se:kth:diva-313639
\item Anil, N., Jain, T., Pine, D., Wenzel, M., Victor, Y., \& Parente, J.  (2022, April 13). Designing a DDD-oriented microservice. Microsoft
  Docs. https://docs.microsoft.com/enus/dotnet/architecture/microservices/microservice-ddd-cqrs-patterns/ddd-oriented-microservice
\item Sherer, T., Petersen, T., Cooper, C., Peterson, N., Boeglin, A.,
  Price, E., Kshirsagar, D., Buck, A.,\& Furbush, K. (2022, April 6).
  Identify microservice boundaries - Azure Architecture Center.
  Microsoft Docs.
\end{enumerate} https://docs.microsoft.com/en-us/azure/architecture/microservices/model/microservice-boundaries

\begin{enumerate}
\def\labelenumi{\arabic{enumi}.}
\setcounter{enumi}{14}
\item
  Vettor, R., Pine, D., Warren, G., Coulter, D., Wenzel, M., Schonning,
  N., Nava, A., \& Veloso, M. (2022, April 7). Service-to-service
  communication. Microsoft Docs.  https://docs.microsoft.com/en-us/dotnet/architecture/cloud-native/service-to-service-communication
\item
  Vettor, R., Warren, G., Pine, D., Coulter, D., James, \& Victor, Y.
  (2022, April 16). gRPC. Microsoft Docs.
\end{enumerate} https://docs.microsoft.com/en-us/dotnet/architecture/cloud-native/grpc

\begin{enumerate}
\def\labelenumi{\arabic{enumi}.}
\setcounter{enumi}{16}
\item
  Fowler,M.(2015,May).bliki:https://martinfowler.com/bliki/MicroservicePremium.html\textbackslash{}
\item
  E. Evans, Domain-Driven Design: Tackling Complexity in the Heart of
  Software, Addison-Wesley, 2003.
\item
  H. VURAL and M. KOYUNCU, ``Does Domain-Driven Design Lead to Finding
  the Optimal Modularity of a Microservice?,'' Does Domain-Driven Design
  Lead to Finding the Optimal Modularity of a Microservice?, Feb. 22,
  2021.
\item
  Evans, E., \& Evans, E. J. (2004). Domain-driven design: tackling
  complexity in the heart of software. Addison-Wesley Professional.
\item
  Levcovitz, A., Terra, R., \& Valente, M. T. (2016). Towards a
  technique for extracting microservices from monolithic enterprise
  systems. arXiv preprint arXiv:1605.03175.
\item
  Al-Debagy, O., \& Martinek, P. (2018, November). A comparative review
  of microservices and monolithic architectures.
\item
  Velepucha, V., \& Flores, P. (2021, March). Monoliths to
  microservices-migration problems and challenges: A sms. In~2021 Second
  International Conference on Information Systems and Software
  Technologies (ICI2ST)~(pp. 135-142). IEEE.
\item
  Gos, K., \& Zabierowski, W. (2020, April). The comparison of
  microservice and monolithic architecture. In 2020 IEEE XVIth
  International Conference on the Perspective Technologies and Methods
  in MEMS Design (MEMSTECH) (pp. 150-153). IEEE.
\item
  De Lauretis, L. (2019, October). From monolithic architecture to
  microservices architecture. In 2019 IEEE International Symposium on
  Software Reliability Engineering Workshops (ISSREW) (pp. 93-96). IEEE.
\item
  {[}4{]}. Kuryazov, D., Jabborov, D., \& Khujamuratov, B. (2020,
  October). Towards decomposing monolithic applications into
  microservices. In 2020 IEEE 14th International Conference on
  Application of Information and Communication Technologies (AICT) (pp.
  1-4). IEEE.
\item
  {[}5{]}. Ivanov, N., \& Tasheva, A. (2021, September). A Hot
  Decomposition Procedure: Operational Monolith System to Microservices.
  In 2021 International Conference Automatics and Informatics (ICAI)
  (pp. 182-187). IEEE.
\item
  {[}6{]} Zaki, J., Islam, S. R., Alghamdi, N. S., Abdullah-Al-Wadud,
  M., \& Kwak, K. S. (2022). Introducing cloud-assisted
  micro-service-based software development framework for healthcare
  systems. IEEE Access, 10, 33332-33348.
\item
  {[}7{]} Haugeland, S. G., Nguyen, P. H., Song, H., \& Chauvel, F.
  (2021, September). Migrating monoliths to microservices-based
  customizable multi-tenant cloud-native apps. In 2021 47th Euromicro
  Conference on Software Engineering and Advanced Applications (SEAA)
  (pp. 170-177). IEEE.
\item
  {[}8{]} Gonçalves, N., Faustino, D., Silva, A. R., \& Portela, M.
  (2021, March). Monolith modularization towards microservices:
  Refactoring and performance trade-offs. In~2021 IEEE 18th
  International Conference on Software Architecture Companion
  (ICSA-C)~(pp. 1-8). IEEE.
\item
  {[}9{]} Santos, N., \& Silva, A. R. (2020, March). A complexity metric
  for microservices architecture migration. In 2020 IEEE international
  conference on software architecture (ICSA) (pp. 169-178). IEEE.
\item
  {[}10{]} Blinowski, G., Ojdowska, A., \& Przybyłek, A. (2022).
  Monolithic vs. microservice architecture: A performance and
  scalability evaluation. IEEE Access, 10, 20357-20374.
\end{enumerate}

\end{document}